\documentstyle[12pt]{article}

\begin{document}

\bigskip \medskip \begin{center} \large {\bf The Spectral Problem for
the ${\bf q}$-Knizhnik-Zamolodchikov Equation} \\

\bigskip \bigskip \normalsize

Peter G. O. Freund\footnote{Work supported in part by the NSF:
PHY-91-23780} and Anton V. Zabrodin\footnote{Permanent address:
Institute of Chemical Physics, Kosygina Str. 4, SU-117334, Moscow,
Russia }\\ {\it Enrico Fermi Institute, Department of Physics \\ and
Mathematical Disciplines Center \\ University of Chicago, Chicago, IL
60637} \\

\end{center}

\bigskip \centerline{\bf ABSTRACT} \begin{quote} We analyze the
spectral problem for the $q$\--Knizh\-nik\--Zamo\-lod\-chi\-kov
equations for $U_q(\widehat{sl_2}) (0 < q \leq 1)$ at level zero.  The
case of 2\--point functions in the fundamental representation is
studied in detail.  The scattering states are found explicitly in
terms of continuous $q$\--Jacobi polynomials.  The corresponding
$S$\--matrix is shown to coincide, up to a trivial factor, with the
kink\--antikink $S$\--matrix in the spin\--${1 \over 2}$ XXZ
antiferromagnet.

\end{quote}

In conformal field theory, the matrix elements of products of vertex
operators between suitable vacuum states obey the
Knizhnik\--Zamolodchikov (KZ) equations [1].  These are first order
differential equations akin to the Dirac and Bargmann\--Wigner
equations [2].  Recently Frenkel and Reshetikhin [3] have derived a
$q$\--analogue of the KZ equation for {\em quantum} affine algebras.
This is a considerable generalization and these $q$\--KZ equations are
no longer differential, but rather difference equations.

In this paper we study in detail the spectral problem for the $q$\--KZ
equation of $U_q(\widehat{sl_2})$ in the fundamental representation at
level zero.  The scattering states are explicitly found in terms of
continuous $q$\--Jacobi polynomials.  From the known asymptotics of
these polynomials we derive an $S$\--matrix which coincides, up to a trivial
factor, with the $S$\--matrix for kink\--antikink scattering in the
spin ${1 \over 2}$ XXZ antiferromagnet.  Even in the ``classical''
limit $q \rightarrow 1$, this {\em spectral} problem yields a
non\--trivial $S$\--matrix, corresponding of course to the XXX model.
This is quite different from the monodromy problem, which yields a
non\--trivial, spectral parameter dependent $S$\--matrix at the
($q$\--deformed) quantum level, but {\em not} at the classical $(q
\rightarrow 1)$ level.  The spectral problem is dual, as it were, to
the monodromy problem.

As will be clear below, the present work is closely related to our
earlier work [4-7] on Macdonald and Hall\--Littlewood\--Kerov
polynomials as zonal spherical functions for certain
quantum\--symmetric spaces.  Such spherical functions are
eigenfunctions of a {\em second} order difference operator, the
analogue of the Laplace operator.  Zonal spherical functions carry
information only about a particular component of the physical
excitations.  For the XXX and XXZ models the excitations are known to
be spin\--${1 \over 2}$ kinks [8,9].  To obtain the {\em full}
kink\--antikink $S$\--matrix one must consider the spectral problem
for the (``Dirac\--like'') $q$\--KZ operator, rather than for a scalar
Laplace\--like operator.

To fix ideas, consider first the classical $(q = 1)$ KZ\--equation for
$\widehat{sl_2}$ in the fundamental representation for level zero.
Here everything is very simple and the spectral problem is readily
solved.  Call $\Phi (x_i)$ the vertex operators and define normal
ordering relative to a Borel polarization.  The matrix element of
interest $$ \Psi (x_i) = \langle \Omega ' | \Phi (x_2) \Phi (x_1) |
\Omega \rangle \eqno(1) $$ depends only on one variable, which in an
{\em additive} parametrization can be chosen as $$ x = {1 \over 2}
(x_1 - x_2) ~.  \eqno(2) $$ The matrix element $\Psi (x_i)$ is then a
${\bf C}^2 \otimes {\bf C}^2$ valued function $\Psi(x)$ of the
variable $x$.  This $\Psi (x)$ obeys the KZ equation, which, for
$\widehat{sl_2}$ at level zero and in the fundamental representation,
takes the form $$ 2 \frac{d \Psi (x)}{dx} = (r_{12} (x) + \pi_1 (H))
\Psi (x) ~, \eqno(3) $$ where $r_{12} (x)$ is the familiar
trigonometric solution of the {\em classical} Yang\--Baxter equation
$$ \begin{array}{l} r_{12} (x) = (\coth z) [ E \otimes F + F \otimes E
+ {1 \over 2}
 H \otimes H - {1 \over 2} {\bf 1} \otimes {\bf 1} ] - E \otimes F + F
\otimes E \\ \\ \pi_1 (H) = i \lambda H \otimes {\bf 1}, ~~ E = \left
( \begin{array}{c c} 0 &1 \\ 0 & 0 \end{array} \right ) , F = \left (
\begin{array}{c c} 0 & 0 \\ 1 & 0 \end{array} \right ) , H = \left (
\begin{array}{c c} 1 &0 \\ 0 & -1 \end{array} \right ) ~, \end{array}
\eqno(4) $$ with $\lambda$ the (real) parameter of our spectral
problem.  The prefactor 2 on the left hand side of eq.~(3) stands for
the combination $k+g$, with the level $k$ already set at $k=0$ and
with $g$, the dual Coxeter number set by $sl_2$ at $g=2$.  The choice
of the term proportional to the $4 \times 4$ unit matrix in the
expression of $r_{12} (x)$ (the last term in the square bracket in the
first equation (4)) is irrelevant as far as the Yang\--Baxter equation
is concerned, but considerably simplifies the argument.  With respect
to a basis $v_a \otimes v_b , a, b = \pm$ of ${\bf C}^2 \otimes {\bf
C}^2$, we can expand $$ \Psi (x) = a (x) v_+ \otimes v_+ + f (x) v_+
\otimes v_- + g (x) v_- \otimes v_+ +
 b (x) v_- \otimes v_- \eqno(5a) $$ In components the KZ equation (3)
then becomes $$
 {{d} \over {dx}} a (x) = i \lambda a (x), ~~
 {{d} \over {dx}} b (x) = - i \lambda b (x) \eqno(5b) $$ $$ \left (
\frac{d}{dx} + {1 \over 2} \coth x - {1 \over 2} \sigma_1 \coth x
\right ) \psi (x) = i ( \lambda \sigma_3 - {1 \over 2} \sigma_2 ) \psi
(x), ~~ \psi (x) = \left ( \begin{array}{c} f (x) \\ g (x) \end{array}
\right ) \eqno(5c) $$ The nonzero weight com\-po\-nents $a(x)$ and
$b(x)$ de\-couple and are trivi\-al \\ (eqs.~(5b)).  Hence\-forth we
ignore them.  The interesting equation (5c) involves the zero weight
sector.  This equation is similar to a Dirac equation with our
spectral parameter $\lambda$ playing the role of mass in a
``$\gamma_5$\--type'' mass term.  It is convenient to introduce the
triplet and singlet combinations $$ F^+ (x) = f(x) + g (x), ~~ F^- (x)
= f(x) - g (x) ~, \eqno(6) $$ which obey the first order differential
equations $$ F^{- \prime } + (\coth x ) F^- = (i \lambda - {1 \over 2}
) F^+, ~~ F^{+ \prime } = (i \lambda + {1 \over 2} ) F^- ~.  \eqno(7)
$$ Here prime stands for derivative with respect to $x$.  The
equations (7) lead to the decoupled system of second order
differential equations $$ \left ( {{d^2} \over {dx^2}} + \coth x {{d}
\over {dx} }\right ) F^+ (x) = - \left ( \lambda^2 + {1 \over 4}
\right ) F^+ (x) $$ $$ \left ( {{d^2} \over {dx^2}} + \coth x {d \over
{dx}}- {{1} \over {\sinh^2 x}} \right ) F^- (x) = - \left ( \lambda^2
+ {1 \over 4} \right ) F^- (x) ~.  \eqno(8) $$ The first of these
equations is a special case of Matsuo's theorem [10].  The connection
with the theory of spherical functions is now immediate.  The
Laplace\--Beltrami operator on the real hyperbolic plane $H$ is $$
\Delta = \frac{\partial^2}{\partial x^2} + \coth x
\frac{\partial}{\partial x} + \frac{1}{\sinh^2 x}
\frac{\partial^2}{\partial \theta^2} ~.  \eqno(9) $$ Its
eigenfunctions $G_m (x ,\theta )$ with appropriate boundary conditions
(finiteness at the origin) are the spherical functions $$ \Delta G_m
(x , \theta ) = - \left ( \lambda^2 + {1 \over 4} \right ) G_m (x ,
\theta ) \eqno(10) $$ where $m$ is the angular momentum $$ - i
\frac{\partial}{\partial \theta} G_m (x , \theta ) = m G_m (x , \theta
) \eqno(11) $$ Zonal spherical functions correspond to $m=0$, tesseral
ones to integer $m \neq 0$.  It is readily seen that the second order
equations (8), derived from the KZ equations, are precisely the radial
parts of the equations (10) with $m=0$ for $F^+(x)$ (which is thus
zonal) and $m=1$ for $F^- (x)$ (which is thus tesseral).  We impose on
$F^\pm$ the same boundary conditions as on the spherical functions
whose equations they obey.  Then, the extraction of an $S$\--matrix
from the solutions $F^\pm$ of these equations becomes a familiar
story.  In terms of the Harish\--Chandra $c$\--function on $H$ [11] $$
c(\lambda ) = \pi^{1/2} \frac{\Gamma ( i \lambda )}{\Gamma (i \lambda
+ {1 \over 2})} ~, \eqno(12) $$ the $S$-matrix elements are $$ S_+
(\lambda ) = \frac{c(\lambda )}{c ( - \lambda )}, ~~ S_- (\lambda ) =
B ( \lambda ) S_+ (\lambda ),~~ B ( \lambda ) = \frac{1 + i2 \lambda
}{1 - i 2 \lambda} ~.  \eqno(13) $$ The singlet $S_-$ and triplet
$S_+$ $S$-matrix elements differ only by the Blaschke\--CDD\--pole
factor $B ( \lambda )$.  Together $S_+ ( \lambda)$ and $S_- ( \lambda
)$ describe kink\--antikink scattering in the Heisenberg XXX model.
The eigenvalues of the corresponding $S$-matrix are precisely $S_+ (
\lambda)$ and $-S_- ( \lambda)$ (the minus sign will be discussed
below, following eq. (32)).
In what follows, we shall show that all this generalizes to the $q$\--KZ
equation, in the ($q$\--deformed) quantum case.
Before doing so, let us still comment on two features of the classical $(
q=1)$ case.

First, as far as $sl_2$ representations are concerned, $\Psi (\lambda )$
is valued in the tensor product of two 2\--dimensional representations
and as such decomposes into one singlet and three triplet components.
Yet, as we saw, the two non\--vanishing weight components of the triplet
decouple and what is left is a two\--dimensional ``spin\--1/2''\--like
system on which the Dirac\--like equation (5c) is placed.
It would be interesting to further study this metamorphosis.

Second, the steps leading from the coupled first order system (7), to the
decoupled second order system (8) are completely parallel with going from
the coupled first order Dirac equations to the uncoupled eigenfunctions of
the Laplace operator.
In fact, as we saw, we do obtain the eigenfunctions of a Laplace operator
on the real hyperbolic plane.
It will be worth keeping this in mind when we repeat the same steps at
the quantum level.
We now come to the case of the $q$\--KZ equations.

Let us start by recalling some facts about $q$\--KZ equations for
2\--point functions in $U_q ( \widehat{sl_2})$\--case $(0 < q < 1)$.
We use the notations of [12], in a slightly modified form.
Consider a correlation function of two $q$\--vertex operators
$$
\Psi (z_1 , z_2 ) = \langle \Omega ' | \Phi (z_2) \Phi (z_1 ) | \Omega
\rangle \in V \otimes V
\eqno(14)
$$
where $V \cong {\bf C}^2$ is a linear space on which  now 2\--dimensional
representations of $U_q ( \widehat{sl_2})$ act.
We fix a basis $\{v_+ , v_- \}$ in $V$.
With a proper definition of $q$\--vertex operators, $\Psi (z_1 , z_2 )$
depends only on $z_1/z_2$
(we have now switched to a multiplicative parametrization),
so we consider the function
$$
\Psi (z) = \langle \Omega ' | \Phi (z^{-1}) \Phi (z) | \Omega \rangle ~.
\eqno(15)
$$

For level $k$ $q$\--vertex operators, $\Psi (z)$ satisfies the
first order $q$\--KZ
difference equation:
$$
\Psi (q^{k+2} z) = \rho (z) (q^{- \phi} \otimes 1 ) R (z) \Psi (z)
\eqno(16)
$$
where the $R$\--matrix $R(z)$ is defined by explicit action in $V \otimes
V$ as follows:
$$
R(z) v_\pm \otimes v_\pm = v_\pm \otimes v_\pm
\eqno(17a)
$$
$$
R(z)v_+ \otimes v_- =
{{q(1-z^2)} \over {1-q^2z^2}} v_+ \otimes v_- +
{{(1-q^2)z^2} \over {1-q^2 z^2}} v_- \otimes v_+
$$
$$
R(z)v_- \otimes v_+ =
{{1-q^2} \over {1-q^2z^2}} v_+ \otimes v_- +
{{q(1-z^2)} \over {1-q^2z^2}} v_- \otimes v_+
\eqno(17b)
$$
The operator $q^{- \phi}$ acts on the basis vectors by multiplication:
$$
q^{- \phi} v_\pm = q^{{\mp}2i \lambda} v_\pm
\eqno(18)
$$
where $\lambda$ is a spectral parameter.
In the $q$\--KZ equations considered in [3, 12] $\lambda$ takes a
particular value depending on the choice of vacuum states $\Omega, \Omega
'$ in (15).
When we are interested in the spectral problem for the difference
operator in (16) (rather than monodromy properties of the solutions)
$\lambda$ plays the role of spectral parameter (this becomes evident from
eq.~(23) below).
We have introduced $i$ in (18) (just like in eq.~(4) in the classical
case) so that the continuous spectrum will correspond to real values of
$\lambda$.

Finally, $\rho(z)$ in (16) is the scalar multiplier defined by [12]:
$$
\rho(z) = q^{- 1/2}
\frac{(q^2z^2; q^4)_\infty^2}{(z^2; q^4)_\infty(q^4z^2; q^4)_\infty}
\eqno(19)
$$
where the standard notation [13]
$$
(z;q)_n = \prod_{j=0}^{n-1} (1-zq^j); ~~
(z; q)_\infty = \lim_{n \rightarrow \infty} (z; q)_n
\eqno(20)
$$
is used.
It is shown in [3] that this multiplier comes from restriction of the
universal $R$\--matrix for $U_q (\widehat{sl_2})$ to the tensor
product of two 2\--dimensional representations.
Though crucial in the monodromy problem, $\rho (z)$ is irrelevant for
our purposes here, because one can gauge it away without altering the
spectral properties.

Due to the specific form (17) of the $R$\--matrix, the $v_+ \otimes v_+$
and $v_- \otimes v_-$ components of $\Psi (z)$ decouple, and each of
them obeys a scalar first order difference equation as in the classical case
(5b).
Again, the non\--trivial equations come from the zero\--weight sector of
the $R$\--matrix (17b).
After reduction to the zero\--weight subspace of $V \otimes V$ we obtain
for the two components of
$$
\psi (z)  \equiv f(z)v_+ \otimes v_- +
g(z) v_- \otimes v_+ ~,
\eqno(21)
$$
the following system of difference equations
$$
f(q^{k+2} z) = q^{-2i \lambda}
{{q(1-z^2)} \over {1-q^2z^2}} f(z) + q^{-2i \lambda}
{{1-q^2} \over {1-q^2z^2}} g (z)
$$
$$
g(q^{k+2} z) = q^{2i \lambda}
{{(1-q^2)z^2} \over {1-q^2z^2}} f(z) + q^{2i \lambda}
{{q(1-z^2)} \over {1-q^2z^2}} g (z)
\eqno(22)
$$
{}From now on we restrict ourselves to the case of level zero $(k=0)$.
As we shall see, the spectral problem (6) for this simplest case gives
physical $S$\--matrices for the XXZ model having at the same time a nice
geometrical interpretation in terms of scattering on the quantum group
$SL_q(2, {\bf R})$.

There is a more suggestive form of (22) which resembles the classical
spectral problem (5c).
Calling $R_0(z)$ the zero\--weight part (17b) of the $R$\--matrix and
introducing the diagonal $2 \times 2$ matrix $\Lambda = {\rm diag}
\{q^{2i \lambda}, q^{-2i \lambda} \}$, one can rewrite (22) in the form
$$
T^{-1}R_0 (z) \psi (z) = \Lambda \psi (z)
\eqno(23)
$$
where $T$ is the shift operator:  $T \psi (z) = \psi (q^2z)$ (for $k=0$).
This equation does look like a finite\--difference analog of (5c).

Guided by the classical limit we interpret (23) as the radial part of a
discrete ``Dirac\--like'' equation for a particle on a curved quantum
space.
Let us rewrite (22), (23) in terms of the discrete ``radial coordinate''
$n$ which we assume to be a non\--negative integer.
Setting $z=q^{2n+1}$ and calling $f(q^{2n+1}) = f_n$, $g(q^{2n+1})
=g_n$, we obtain the following system of recurrence relations
$$
\left (
\begin{array}{c}
f_{n+1} \\
g_{n+1}
\end{array}
\right )
= (1-q^{4n+4})^{-1}
\left (
\begin{array}{c c}
q^{-2i \lambda +1}(1-q^{4n+2}) & q^{-2i \lambda}(1-q^2) \\
q^{2i \lambda} (1-q^2)q^{4n+2} & q^{2i \lambda + 1} (1-q^{4n+2} )
\end{array}
\right )
\left (
\begin{array}{c}
f_n \\
g_n
\end{array}
\right )
\eqno(24)
$$
By a straightforward but somewhat lengthy calculation it can be shown
that the linear combinations
$$
F_n^\pm = f_n \pm q^{-2i \lambda} g_n
\eqno(25)
$$
obey second order recurrence relations of the form
$$
\frac{1-q^{4n+4}}{1-q^{4n+2}} F_{n+1}^\pm + q^2
\frac{1-q^{4n-4}}{1-q^{4n-2}} F_{n-1}^\pm =
$$
$$
\mp
\frac{(1-q^2)(1-q^4)q^{4n-2}}{(1-q^{4n+2})(1-q^{4n-2})}
F_n^\pm +q
\left ( q^{2i \lambda} + q^{-2i \lambda} \right ) F_n^\pm
\eqno(26)
$$
(Equations (25) and (26) are the quantum counterparts of the
``classical'' equations (6) and (8)).

A natural boundary condition is again the finiteness of $F_n^\pm$ at $n=0$.
Note that due to the specific form of the coefficients in (26), we don't
need to fix values of $F_n^\pm$ at {\em two} points (say, $n=0$ and
$n=1$),
which would be the usual thing to do for second\--order recurrence
relations.
In the case at hand the regularity at $n=0$ already determines the
solutions $F_n^\pm$ up to arbitrary constant factors $a^\pm$.
Note also that, had we included the factor $\rho(z)$ (5) in the
$R$\--matrix, its effect would be to force $F_n^\pm$ to be zero at
negative $n$ because the zeros of $\rho (z)$ lie just at the points
$q^{2n+1} , ~ n < 0$.

Notice that the eqs.~(26) are the recurrence relations for certain
$q$\--Jacobi polynomials [13] (with these very boundary conditions).
Specifically,
$$
F_n^+ = a^+ \cdot P_{n-1}^{(0,1)} \left ( {1 \over 2} (q^{2i \lambda} +
q^{-2 i \lambda}  ); q^2 \right )
\eqno(27a)
$$
$$
F_n^- = a^- \cdot (q^{-2n} - q^{2n}) P_{n-1}^{(1,0)} \left (
{1 \over 2} (q^{2 i \lambda} + q^{-2 i \lambda} ) ; q^2 \right )
\eqno(27b)
$$
Here $P_n^{( \alpha , \beta )} (x;q^2 )$ are the continuous $q$\--Jacobi
polynomials defined by the following $q$\--hypergeometric series [13]:
$$
P_n^{(\alpha , \beta )} (x;q) = _4\phi_3
\left (
\begin{array}{l r}
q^{-n} , q^{n+ \alpha + \beta +1} , q^{1/2} e^{i \theta} ,
q^{1/2} e^{-i \theta} & \\
    & ; ~~ q, q \\
q^{\alpha + 1} , -q^{\beta+1} , - q & \\
\end{array}
\right ) =
$$
$$
= \sum_{k=0}^n
\frac{(q^{-n} ; q)_k (q^{n+ \alpha + \beta + 1} ;q)_k (q^{1/2} e^{i
\theta} ; q)_k (q^{1/2} e^{-i \theta} ; q)_k }{(q; q)_k (q^{\alpha+1};
q)_k (-q^{\beta+1}; q)_k (-q; q)_k } q^k
\eqno(28)
$$
where $x = \cos \theta$
(our normalization differs from that of [13]).
These are polynomials in $x$ of degree $n$.
It is known that for $\alpha \geq - {1 \over 2} , \beta \geq - {1 \over
2}$ they form an orthogonal family with a positive measure on the
interval $[-1, 1]$. As in the classical case, the $q$\--Jacobi polynomials
are known to provide (for some values of $\alpha , \beta$) the full set of
spherical harmonics on the quantum group $SL_q(2)$ [14].
Recalling that in this case $\alpha = |m-n|, \beta = |m+n|$ where $m(n)$
is the number of the left (right) $SO(2)$\--harmonics, we see that the
values of $\alpha , \beta$ in (27) correspond to spinorial harmonics $(
{1 \over 2} , {1 \over 2})$ and $({1 \over 2} , - {1 \over 2})$.
This is in line with the Dirac\--Bargmann\--Wigner analogy.

For real values of $\lambda$ in the Brillouin zone $- \frac{\pi}{2 \log
q} < \lambda \leq \frac{\pi}{2 \log q}$ the wave functions  (27) are the
scattering states for the $q$\--KZ operator (l.h.s. of eq.~(23)).
To see this it is necessary to find the asymptotics of $q$\--Jacobi
polynomials at large $n$.
Fortunately, this is known [13].
Using this result we obtain:
$$
F_n^\pm |_{\scriptstyle{n \rightarrow \infty}}\sim
a^\pm q^n \left ( q^{2 i \lambda n} c_\pm (- \lambda ) + q^{-2i \lambda n}
c_\pm (\lambda ) \right )
\eqno(29)
$$
where (up to a non\--essential constant)
$$
c_\pm (\lambda ) =
\frac{q^{2 i \lambda}}{1 \pm q^{2 i \lambda +1}}
\frac{\Gamma_{q^4} ( i \lambda)}{\Gamma_{q^4}(i \lambda + {1 \over 2})} ~,
\eqno(30)
$$
and  the $q$-gamma function is defined by [13]
$$
\Gamma_q(x) =
\frac{(q; q)_\infty}{(q^x ; q)_\infty} (1-q)^{1-x}
\eqno(31)
$$
The expression (29) does indeed look like a superposition of incoming and
outgoing waves, so we can define
$$
S_\pm (\lambda) =
\frac{c_\pm (\lambda )}{c_\pm (- \lambda )}
\eqno(32)
$$
which are the eigenvalues of the $2 \times 2$ $S$\--matrix
corresponding to the $q$\--KZ equation (23).Note that in the classical
limit $(q \rightarrow 1)$, the eqs.~(32) do indeed reduce to the classical
equations (13).
It is readily shown that the $2 \times 2$ $S$\--matrix with eigenvalues (32)
is proportional to $\Lambda^{-1}\sigma_{1}R_0 (q^{2 i \lambda})$ (with the
same matrix $\Lambda$ as in eq. (23)) with the scalar infinite product factor
ensuring unitarity and with
crossing symmetry given by the $q\--\Gamma$\--functions in (30).
So the local (bare) ``$S$\--matrix'' $R_0(z)$in (23) (connecting two
neighboring sites) reproduces itself in the global scattering, getting
``dressed'' by the scalar infinite product factor.

The most interesting feature of all this is that this $S$\--matrix
actually coincides (up to the trivial matrix factor $\Lambda^{-1} \sigma_1$,
which commutes with $R_0(q^{2i \lambda} )$) with the kink\--antikink
scattering matrix in the
spin-$ 1 \over 2$ XXZ antiferromagnet (see, for example eq.~(6.18) in [12]);
$\pm S_\pm$ given by (32) are just its eigenvalues (corresponding to
scattering with a given parity). The role of the extra matrix factor
$ \Lambda^{-1}\sigma_1$
is to change the sign of $S_-$. In the classical limit, this factor reduces
to the constant matrix $ \sigma_1$ and produces the minus sign noted there.

We conclude with a few remarks.

It is interesting that in the $q$\--deformed case the choice (25) of
linear combinations leading to a reasonable pair of decoupled
second\--order equations is {\em not unique}.
The other possibility is $\tilde{F}_n^\pm = f_n \pm q^{-4i \lambda \pm
1}g_n$ (the  $q \rightarrow 1$ limit is the same!).
These $\tilde{F}_n^+$ can be identified with the
Rogers\--Askey\--Ismail polynomials [13] (Macdonald polynomials for root
system $A_1$ [15]).
A similar result was obtained by Cherednik [16].
Remarkably, the asymptotic form of $\tilde{F}_n^+$ gives yet another
eigenvalue of the full XXZ $S$\--matrix corresponding to {\em
kink\--kink} (or antikink\--antikink) scattering (this is just the result
obtained in our earlier papers [4-7] by considering the scalar spectral
problem on a quantum hyperbolic plane).
The remaining (``tesseral'') component $\tilde{F}_n^-$, similarly to
what happened in the case of eq.~(27b) and of the $F^-$ solution to
eq.~(8),
can also be expressed in terms of Rogers\--Askey\--Ismail polynomials.
The boundary conditions for $\tilde{F}_n^\pm$ differ from those for
$F_n^\pm$.

If we take the classical limit $q \rightarrow 1$, for fixed $\varphi = 2
\lambda
\log q$ (so that $q^{2 i \lambda} \rightarrow e^{i \varphi})$ the
$q$\--Jacobi polynomials $P_n^{(\alpha , \beta )} ({1 \over 2} (q^{2 i
\lambda} + q^{-2i \lambda}); q^{2})$ go to the classical Jacobi polynomials
$P_n^{(\alpha , \beta )} (\cos \varphi )$ up to normalization.
To achieve  agreement with the $S$\--matrix (13) we need another
classical limit!
This is $q \rightarrow 1$ and $n \rightarrow \infty$  for fixed $x \sim n \log
q$ and $\lambda$.
This limiting procedure gives the right XXX $S$\--matrix (13).
In this case the $q$\--Jacobi polynomials go to Legendre (or
Gegenbauer) functions $P_{- \alpha - 1/2 + i \lambda} (\cosh x)$ and {\em
not} to Jacobi functions (as one would expect).
This can be easily seen from (28) and  was already pointed out by
Koornwinder [17].

In this paper we only dealt with level zero in eq.~(22).
The scattering problem for non\--zero $k$ can be considered along the
same lines.
A generalization to $q$\--KZ equations for multipoint functions is also
of interest.
It should correspond to the multiparticle scattering in XXZ spin chain
which is known to factorize.
We hope to discuss these questions elsewhere.

The last comment is about monodromy versus the scattering problem.
Our results suggest that they are in some sense dual to each other.
The usual setting of the monodromy (or connection) problem for $q$\--KZ
equations is quite the opposite to what we have done:  $\lambda$ is fixed
once and for all and one compares the solutions regular at $z=0$
with those regular at $z
= \infty $ (in the variable $n$ this corresponds to $n = - \infty$ and $n =
\infty$).
Then the XXZ $S$\--matrix appears as the ratio of two special
solutions and it is now a function of $z$, not $\lambda$.
This indicates a remarkable duality between the coordinate variable and
the spectral parameter.

\bigskip

We are grateful to A. Gorsky, S. Shatashvili and P. Wiegmann  for
discussions and to I. Cherednik and T. Koorwinder for sending us their
papers.
One of us (A.Z.) wishes to thank Professor J.~Peter May for the hospitality
of the Mathematical Disciplines Center at the University of Chicago.

\bigskip
\leftline{\bf REFERENCES}

\medskip
%1
\noindent
[1] V.~G.~Knizhnik and A.~B.~Zamolodchikov, Nuclear Physics {\bf B247}
(1984) 83.

\smallskip
%2
\noindent
[2] V.~Bargmann and  E.~P.~Wigner, Proc. Nat. Acad.Sci. USA {\bf 34} (1946)
211.

\smallskip
%3
\noindent
[3] I.~B.~Frenkel and N.~Yu.~Reshetikhin, Comm. Math. Phys. {\bf 146}
(1992) 1.

\smallskip

%4
\noindent
[4] A.~V.~Zabrodin, Mod. Phys. Lett. {\bf A7} (1992) 441.

\smallskip

%5
\noindent
[5] P.~G.~O.~Freund and A.~V.~Zabrodin, Comm. Math. Phys. {\bf 147} (1992) 277.

\smallskip

%6
\noindent
[6] P.~G.~O.~Freund and A.~V.~Zabrodin, Phys. Lett. {\bf B284} (1992) 283.

\smallskip

%7
\noindent
[7] P.~G.~O.~Freund and A.~V.~~Zabrodin, Phys. Lett. {\bf B294} (1992) 347.

\smallskip

%8
\noindent
[8] L.~D.~Faddeev and A.~L.~Takhtadjan, Zapiski Nauchn. Sem. LOMI, {\bf 109}
(1981) 134.

\smallskip

%9
\noindent
[9] A.~N.~Kirillov and N.~Yu.~Reshetikhin, J.~Phys.~{\bf A20} (1987) 1565.

\smallskip

%10
\noindent
[10] A.~Matsuo, Invent. Math. {\bf 110} (1992) 95.

\smallskip

%11
\noindent
[11] S.~Helgason, {\em Groups and Geometric Analysis} Academic Press,
Orlando 1984.

\smallskip

%12
\noindent
[12] B.~Davies, O.~Foda, M.~Jimbo, T.~Miwa and A. Nakayashiki, Comm.
Math. Phys. {\bf 151} (1993) 89.

\smallskip

%13
\noindent
[13] G.~Gasper and M.~Rahman, {\em Basic Hypergeometric Series}, Cambridge:
Cambridge University Press 1990.

\smallskip
%14
\noindent
[14] M.~Noumi and K.~Mimachi, Duke Math. J. {\bf 63} (1991) 65.

\smallskip

%15
\noindent
[15] I.~Macdonald, Queen Mary College preprint 1989.

\smallskip

%16
\noindent
[16] I.~Cherednik, Comm. Math. Phys. {\bf 150} (1992) 109.

\smallskip

%17
\noindent
[17] T.~Koornwinder, Journal of Mathematical Analysis and Applications
{\bf 148} (1990) 44.

\end{document}